\documentclass[useAMS]{mn2e}
\usepackage{times}

\newtheorem{rule-def}[theorem]{Rule}

\newcommand{\be}{\begin{equation}}
\newcommand{\ee}{\end{equation}}
\newcommand{\bea}{\begin{eqnarray}}
\newcommand{\eea}{\end{eqnarray}}

\begin{document}

\title[Dark energy models...]{\bf Dark energy models with time-dependent
  gravitational constant}

\author[Ray \& Mukhopadhyay]{Saibal Ray$^{1,2}$ \& Utpal
Mukhopadhyay$^{3}$\\ $^{1}$Department of Physics, Barasat
Government College, North 24 Parganas, Kolkata 700 124, West
Bengal, India \\ $^{2}$Inter-University Centre for Astronomy and
Astrophysics, PO Box 4, Pune 411 007, India; e-mail:
saibal@iucaa.ernet.in\\ $^{3}$Satyabharati Vidyapith, North 24
Parganas, Kolkata 700 126, West Bengal, India.}

\maketitle

\begin{abstract}
{Two phenomenological models of $\Lambda$, viz. $\Lambda \sim
(\dot a/a)^2$ and $\Lambda \sim \ddot a/a$ are studied under the
assumption that $G$ is a time-variable parameter. Both models show
that $G$ is inversely proportional to time as suggested earlier by
others including Dirac. The models considered here can be matched
with observational results by properly tuning the parameters of
the models. Our analysis shows that $\Lambda \sim \ddot a/a$ model
corresponds to a repulsive situation and hence correlates with the
present status of the accelerating Universe. The other model
$\Lambda \sim (\dot a/a)^2$ is, in general, attractive in nature.
Moreover, it is seen that due to the combined effect of
time-variable $\Lambda$ and $G$ the Universe evolved with
acceleration as well as deceleration. This later one indicates a
Big Crunch.}

\end{abstract}


\section{Introduction}
 The idea of variability of $G$ originated with the work of Dirac
(1937) who for the first time drew the attention of the scientific
community
 towards the possibility of a time-dependent gravitational constant in
 the context of a cosmological model. Afterwards, cosmological
 theories like Brans-Dicke theory (1961), Hoyle-Narlikar theory (1972) and
 the theory of Dirac (1973) himself supported the idea of a
 time-decreasing gravitational constant. Since, in the classical form
 of the general relativity theory $G$ should remain constant, then the
 theories concerned with the variation of $G$ must be, to some extent,
 consistent with relativity theory. The last three theories mentioned
 above are reconcilable with the theory of General Relativity from the viewpoint
 of the perihelion advancement of Mercury and the bending of star
 light. The theory of a expanding Universe also supports the idea of a
 time-dependent ($\dot G/G = \sigma H_0$, where $H_O$ is the Hubble
 parameter and $\sigma$ is a dimensionless parameter depending on the
 gravitational theory as well as on the particular cosmological model)
 gravitational constant (Will, 1987). After the emergence of superstring
 theory, in which $G$ is considered as a dynamical quantity (Marciano 1984), a
 resurrection of the idea of an evolving $G$ occurred. It has been
 shown that a scale-dependent $G$ can represent the dark matter
 (Goldman 1992). Also, there remain some scale-wise discrepancies in the value
 of the Hubble parameter. These discrepancies can be removed if $G$ is
 considered as a variable quantity (Bertolami et al. 1993). Recently,
 \u{S}tefan\u{c}i\'{c} (2004) has
 considered a phantom energy model with time-varying $G$ in which
 exchange of energy and momentum between vacuum and non-relativistic
 matter (or radiation) occur in such a way that the energy-momentum
 tensors $T^{\mu\nu}_\Lambda$ and  $T^{\mu\nu}_m$ are not separately
 conserved, but as a whole $T^{\mu\nu} = T^{\mu\nu}_\Lambda +
 T^{\mu\nu}_m$ is conserved. Variability of $G$ is also supported by
 observational results coming up from Lunar Laser Ranging
 (Turyshev et al. 2003), spinning rate of pulsars (Arzoumanian 1995;
 Kaspi, Taylor \& Ryba 1994; Stairs 2003) Viking Lander (Hellings 1987;
 Reasemberg 1983), distant Type Ia supernova observation (Gaztanaga et al. 2002),
Helioseismological data (Guenther et al. 1998), white dwarf
G117-B15A (Biesiada \& Malec 2004; Benvenuto et al. 2004) etc.

On the other hand, instead of a cosmological constant $\Lambda$,
the recent trend of searching the nature of dark energy, the
driving force for accelerating the Universe, is to select a
$\Lambda$-model of phenomenological character. Recently, the
equivalent relationship of three kinematical models of $\Lambda$,
viz. $\Lambda \sim (\dot a/a)^2$, $\Lambda \sim \ddot a/a$ and
$\Lambda \sim \rho$ have been shown by us (Ray \& Mukhopadhyay
2004) within the framework of constant $G$. Since the idea of
time-dependent $G$ is supported by various theories and
observations, so it is quite natural to investigate the behaviour
of some of the $\Lambda$-models mentioned above when $G$ is a
function of time. This is the motivation behind the present work.

However, it can be stated that invariant property of $\Lambda$
under Lorentz transformation is not satisfied for arbitrary
systems, e.g., material systems and radiation. In this connection
it have shown (Gliner 1965; Majernik 2001) that the energy density
of vacuum represents a scalar function of the four-dimensional
space-time coordinates so that it satisfies the Lorentz symmetry.
We would also like to mention here that Vishwakarma (2001) has
considered a particular Ricci-symmetry under the framework of
general relativity which is the contracted Ricci-collineation
along the fluid flow vector and shows that this symmetry does
demand Lambda to be a function of time (and space, in general).

 In favour of Lambda-decay scenario, irrespective of whether
 they come from extended theories of gravity or phenomenological
 considerations, it is argued that (i) they have been shown to address
 a number of pressing problems in cosmology; (ii) many are independently
 motivated, e.g., by dimensional arguments, or as limiting cases of more
 complicated theories; (iii) most are simple enough that meaningful
 conclusions can be drawn about their viability and (iv) successful
implementation would point toward the eventual Lagrangian
formulation or a more complete theory (Overduin \& Cooperstock
1998).

Under this background the paper is organized as follows: Sec. 2
deals with the solution of the field equations for two different
models (viz. $\Lambda \sim (\dot a/a)^2$, $\Lambda \sim \ddot
a/a$) of $\Lambda$ while amount of variations of $G$, calculated
on the basis of different theories as well as observations are
scrutinized in Sec. 3. Comparison of the present models with
various observational and theoretical results are done in Sec. 4
and conclusions are summarized in Sec. 5.

 \section{Field equations and their solutions}
 The Einstein field
equations
 \bea
  R^{ij} - \frac{1}{2}Rg^{ij} = -8\pi G\left[T^{ij} - \frac{\Lambda}{8\pi G}g^{ij}\right]
\eea
(where $\Lambda$ is the time-dependent cosmological term and
the velocity of light in vacuum, $c = 1$ in relativistic units)
 for the spherically symmetric
Friedmann-Lema{\^i}tre-Robertson-Walker (FLRW) metric
\bea
ds^2 = -dt^2 + a(t)^2\left[\frac{dr^2}{1 - kr^2} + r^2 (d\theta^2 +
sin^2\theta d\phi^2)\right]
\eea
yield respectively the Friedmann and Raychaudhuri
equations in the following forms
\bea
 3\left(\frac{\dot a}{a}\right)^2 = 8\pi G \rho + \Lambda,
\eea
\bea
3\frac{\ddot a}{a} = - 4\pi G(\rho + 3p)+ \Lambda.
\eea
where $a=a(t)$ is the scale factor of the Universe and $k$, the
 curvature constant is taken to be zero for the flat Universe.

 The energy conservation law, as usual, is given by
 \bea
\dot \rho + 3(p + \rho)\frac{\dot a}{a} = 0
 \eea
whereas the time variable-$\Lambda$ and $G$ are governed by the dynamical
condition $\dot \Lambda = - 8\pi \dot G \rho$.

 The barotropic equation of state is
 \bea
  p = \omega \rho
\eea
 where $\omega$ is the equation of state parameter which is, in
 general, a function of time,
scale factor or redshift. However, sometimes it is convenient to
consider $\omega$ as a constant quantity because current
observational data has limited power to distinguish between a time
varying and constant equation of state (Kujat et al. 2002;
Bartelmann et al. 2005). Some useful limits on $\omega$ came from
SNIa data, $-1.67 <\omega < -0.62$ (Knop et al. 2003) whereas
refined values were indicated by the combined SNIa data with CMB
anisotropy and galaxy clustering statistics which is $-1.33 <
\omega < -0.79$ (Tegmark et al. 2004). Since, $\dot a/a = H$ then
from equation (3) we get \bea 4\pi G \rho = \frac{1}{2}(3H^2 -
\Lambda). \eea

 \subsection{Model with $\Lambda \sim  (\dot a/a)^2$}
 By the use of $\Lambda = 3\alpha \left(\frac{\dot a}{a}\right)^2 =
 3\alpha H^2$ , where $\alpha$ is a free parameter, we get from
 equation (7)
\bea 4\pi G \rho = \frac{3(1-\alpha)}{2}H^2. \eea
From equation
(4) using (6), one can obtain
 \bea
3(H^2 + \dot H)= -4\pi G\rho (1+3\omega)+ 3\alpha H^2
 \eea
which by the use of equation (8) takes the form
 \bea
\dot H = -\frac{3(1-\alpha)(1+\omega)}{2}H^2.
 \eea
 Integrating the above equation we
have \bea \frac{\dot a}{a}= H= \frac{2}{3(1-\alpha)(1 +\omega)t}.
 \eea
 Integrating it further one gets
  \bea
  a(t) = C_2t^{\frac{2}{3(1 - \alpha) (1 + \omega)}},
  \eea
where $C_2$ is integration constant.

It is interesting to note that equation (12) is the same expression for
  $a(t)$ as obtained by us (Ray \& Mukhopadhyay 2004) for the same
  $\Lambda$-model with constant $G$.

Again, using equation (12) we get the solution set for the
  matter-energy density, cosmological parameter and gravitational
  parameter respectively as
  \bea
  \rho(t)= C_4 t^{-2/(1- \alpha)}.
 \eea
   \bea
 \Lambda(t) =\frac{4\alpha}{3(1 - \alpha)(1 + \omega)^{2}}t^{2\alpha/(1 - \alpha)}t^{-2},
  \eea
  \bea
  G(t)=\frac{1}{6C_4\pi(1 - \alpha)^{2} (1 + \omega)^{2}}t^{2\alpha/(1 - \alpha)}
\eea
 where $C_4$ is the another
constant of integration and is given by \bea
 C_4 = C_3{C_2}^{-3{1+\alpha}}.
 \eea
Again, we observe that equation (14) is also represents exactly
the
  same expression as in the case of our previous work (Ray \& Mukhopadhyay 2004). This
  means that time variation of $G$ does not affect the scale
  factor as well as the cosmological parameter. From the equation (15) it is easy to obtain
\bea
 \frac{\dot G}{G}= \frac{2\alpha}{1-\alpha}t^{-1}.
\eea

\subsection{Model with $\Lambda \sim \frac{\ddot
a}{a}$}
  Let us put $\Lambda = \beta \frac{\ddot a}{a} = \beta(H^2 + \dot
  H)$, where $\beta$ is a constant. Then, equation (7) becomes
\bea
4\pi G \rho = \frac{3-\beta}{2}H^2 - \frac{\beta}{2}\dot H.
\eea
Using equation (6) we get from equation (4)
\bea
(3-\beta)\dot H + (3-\beta)H^2= - 4\pi G \rho(1 + 3\omega).
\eea
Then by use of the equation (18) the equation (19) transforms to
\bea
\dot H = -\frac{(3-\beta)(1+\omega)}{2-\beta-\beta\omega}H^2.
\eea
Integrating we get \bea \frac{\dot a}{a}= H=
\frac{\beta\omega +\beta -2}{(\beta-3)(1 +\omega)t}.
 \eea
 Integrating again we get
  \bea
  a(t) = C_5t^{\frac{\beta\omega +\beta -2}{(\beta-3)(1 +\omega)}},
  \eea
where $C_5$ is the integration constant.

Using equation (22) we get the general solutions for different
physical parameters as follows: \bea
 \rho(t) = C_6t^{\frac{-3(\beta\omega +\beta -2)}{(\beta-3)(1 +\omega)}},
 \eea
 \bea
\Lambda(t) = \frac{\beta(1+3\omega)(\beta\omega +\beta -2)}{(\beta
-3)^2(1 + w)^2}t^{-2}, \eea
\bea
G(t) = \frac{\beta\omega +\beta
-2}{4\pi C_6(\beta - 3)(1 + \omega)^2} t^{\frac{\beta\omega +\beta
+6\omega}{(\beta-3)(1 +\omega)}}
\eea
where $C_6$ is a constant of
integration.

Equations (22) and (24)are the same expressions for $a(t)$ and
$\Lambda(t)$ as obtained by us (Ray \& Mukhopadhyay 2004). So, for
this model also the scale factor and the cosmological parameter
remain unaffected by time variation of $G$.

\subsection{Comparison with other models}
 From
equation (25), in a similar way as in the previous case, we have
\bea
 \frac{\dot G}{G}= \frac{\beta\omega +\beta +6\omega}{(\beta-3)(1
   +\omega)}
t^{-1}.
\eea
Now, using the expression for $H$ from equation (11),
equation (17) can be written as
 \bea
 \frac{\dot G}{G}= 3\alpha(1 + \omega)H.
\eea
Similarly, putting the expression for $H$ from equation (21),
equation (26) can be written as
 \bea
 \frac{\dot G}{G}= \frac{{\beta}{\omega} + \beta + 6\omega}
{({\beta}{\omega} + \beta - 2)}H. \eea Recently, considering a
time-dependent growing cosmological energy density of the form
\bea \rho_{\Lambda} = \rho_{\Lambda_0}
\left(\frac{a}{a_0}\right)^{-3(1+\eta)} \eea \u{S}tefan\u{c}i\'{c}
(2004) has arrived at the expression for $\dot G/G$ at the present
era as \bea
 \frac{\dot G}{G}= 3(1+\eta)\Omega^0_{\Lambda} H_0.
\eea If we compare equation (27) with equation (21) of
\u{S}tefan\u{c}i\'{c} (2004), then remembering that for the
present era $\Omega^0_{\Lambda} =2/3$, we get for dust
 \bea
\eta= \frac{3\alpha-2}{2}.
 \eea
Similarly, comparing equation (28) with equation (21) of
\u{S}tefan\u{c}i\'{c} (2004), for dust case, we get
 \bea
\eta= \frac{4-\beta}{2(\beta-2)}.
 \eea
From equations (31) and (32) we readily arrive at the relation
\bea 3\alpha = \frac{\beta}{\beta-2}= 2(\eta +1).
 \eea
Equation (33) is interesting, since it inter-relates the parameter
$\eta$ of a phantom energy ($\omega<-1$) model with $\alpha$ and
$\beta$, the parameters of our model with dust ($\omega=0$) case.
Here, for a repulsive $\Lambda$ the constraint on $\alpha$ is
$\alpha>0$.

\section{Present status for the variable-$G$
models}
 The Large Number Hypothesis (LNH) of Dirac prompted him
not to admit the variability of the fundamental constants involved
in atomic physics. Instead he thought of a possible change in $G$
which, in turn, led him to the differential equation (Cetto,
Pe\~{n}a \& Santos 1986)\bea G(t) = k_1 H(t)= k_2
[H(t)]^{3/2}[\rho(t)]^{-1/2} \eea where $k_1$ and $k_2$ are
constants.

From the above equation Dirac obtained, $G(t) \sim t^{-1}$. All the
three variants (early Dirac, additive creation and multiplicative
creation) tells us that $\dot G/G$ should be inversely proportional to
$t$.

According to Brans-Dicke theory (1961), $G$ should vary inversely
with time since according to that theory the scalar field
$\phi(t)$ is time increasing and $G(t) \propto [\phi(t)]^{-1}$.
Dyson (1972,1978) proposed that variation of $G$ should be of the
order of $H$, the Hubble parameter. Since, $H \propto t^{-1}$,
then it is clear that $G$ should decrease as $t^{-1}$.

Coming to the amount of variation of $G$, we find that, relying on
the data provided by three distant quasars of red shift, $z \sim
3.5$ in favour of an increasing fine-structure constant $\alpha$
(Murphy et al. 2002; Webb et al. 2001) and taking the present age
of the Universe as $14$ Gyr, it has been estimated
(Lor\'{e}n-Aguilar et al. 2003) that $\dot G/G \sim + 10^{-15}$
yr$^{-1}$ for Kaluza-Klein and Einstein-Yang-Mills theories
whereas it is of the order of $10^{-13}$yr$^{-1}$ for
Randall-Sundrum theory. The data provided by the binary pulsar PSR
$1913 + 16$, is a very reliable upper bound (Damour et al. 1988),
viz., \bea -(1.10 \pm 1.07) \times 10^{-11} yr^{-1} < \frac{\dot
G}{G}<0 \eea The spinning-down rates of pulsars PSR B$0655 + 64$
(Goldman 1992) and PSR J$2019 + 2425$ (Arzoumanian 1995; Stairs
2003) provide respectively the constraint \bea \left|\frac{\dot
G}{G}\right| \leq (2.2-5.5) \times 10^{-11} yr^{-1} \eea and
 \bea
\left|\frac{\dot G}{G}\right| \leq (1.4-3.2) \times 10^{-11}
yr^{-1}. \eea The range of  $\dot G/G$ provided by the
Helioseismological data (Guenther et al. 1998) is considered as
the best upper bound and is given by \bea -1.60 \times 10^{-12}
yr^{-1} < \frac{\dot G}{G}<0. \eea Data provided by observations
of Type Ia supernova (Riess 1998; Perlmutter 1999) gives the best
upper bound of the variation of $G$ at cosmological ranges as
(Gaztanaga et al. 2002) \bea -10^{-11} yr^{-1} \leq \frac{\dot
G}{G}<0 \quad at\quad z \simeq 0.5. \eea The present cosmological
scenario tells us that we are living in an expanding, flat and
accelerating Universe dominated by dark energy while the remaining
one-third is contributed by matter. So, if we choose
$\Omega_m=0.3$ and $\Omega_\Lambda=0.7$, then the estimated range
of variation of $\dot G/G$ comes out as (Lor\'{e}n-Aguilar 2003)
\bea -1.40 \times 10^{-11} yr^{-1} < \frac{\dot G}{G}<+2.6 \times
10^{-11} yr^{-1}. \eea Very recently, using the data provided by
the pulsating white dwarf star G117-B15A the astereoseismological
bound on $\dot G/G$ is found (Benvenuto et al. 2004) to be \bea
-2.50 \times 10^{-10} yr^{-1} \leq \frac{\dot G}{G}\leq +4.0
\times 10^{-10} yr^{-1} \eea while using the same star Biesiada
and Malec (2004) has inferred that \bea \left|\frac{\dot
G}{G}\right| \leq +4.10 \times 10^{-11} yr^{-1}. \eea On the other
hand, using Big Bang Nucleosynthesis another recent estimate of
variation of $G$ has been obtained (Copi, Davies \& Krauss 2004)
as \bea -4.0 \times 10^{-13} yr^{-1} < \frac{\dot G}{G}<+3.0
\times 10^{-13} yr^{-1}. \eea
 Some other estimates of the probable range of variation of
$G$ can be obtained from various other sources such as Lunar Laser
Ranging (Turyshev et al. 2003), Viking Lander (Hellings 1987;
Resemberg 1983), lunar occultation (Van Flandern 1975), lunar
tidal acceleration (Van Flandern 1975) etc. Some theoretical
models (Blake 1978; Faulkner 1976) also provide estimates of $\dot
G/G$. Various ranges of $\dot G/G$ are listed in Table 1 and 2.

\section{Comparison of present models with
observations}
 Equations (17) and (26) provide us an opportunity
for fitting the $\Lambda \sim (\dot a/a)^2$ and $\Lambda \sim
\ddot a/a$ models respectively with the values of $\dot G/G$
obtained from various sources by proper tuning of $\alpha$ and
$\beta$, the parameters of the two models mentioned above.
Assuming the present age of the Universe as $14$ Gyr, values of
$\alpha$ and $\beta$ corresponding to different values of $\dot
G/G$ are listed in Table 1. From this Table 1 it is evident that
most of the values of $\alpha$ are negative while those of $\beta$
are positive. Now, a negative $\alpha$ means an attractive
$\Lambda$ which does not match with the present status of this
cosmological parameter. However, equation (17) shows that values
of $\dot G/G \sim 10^{-11} yr^{-1}$ can be obtained if we choose
any value of $\alpha>1$. For instance, if we set $\alpha=15$, then
assuming $t_0=14$ Gyr, we get the value of $\dot G/G$ as $-15
\times 10^{-11} yr^{-1}$ which fits well with the value of $\dot
G/G$ for Early Dirac theory (Blake 1978). Coming to the case of
$\Lambda \sim \ddot a/a$ model we find that most of the tuned
values of $\beta$ are positive which corresponds to a repulsive
$\Lambda$. Hence, $\Lambda \sim \ddot a/a$ model can be fitted
more easily with various ranges of $\dot G/G$ than that of
$\Lambda \sim (\dot a/a)^2$ model.

Also, Table 1 shows that although majority of the values of $\dot
G/G$ are negative, but in some cases values of $\dot G/G$ can be
found positive as well. Now, a negative $\dot G/G$ implies a
time-decreasing $G$. This means that by combined effect of a
decreasing $G$ and repulsive $\Lambda$ the Universe will go on
accelerating for ever. On the other hand, a positive $\dot G/G$
means $G$ is growing with time. Since, $\Lambda$ is a
time-decreasing parameter so in future a time-increasing $G$ may
overcome the effect of repulsive $\Lambda$. In that case, the
possibility of a `Big Crunch' cannot be ruled out.

\section{Conclusions}
 In the present work
selecting two different kinematical models of dark energy and
considering the gravitational constant $G$ as a function of time
it has been possible to solve Friedmann and Raychaudhuri equations
for $a(t)$, $\rho(t)$, $\Lambda(t)$ and $G(t)$. Comparing the
expressions for $\dot G/G$ for both the models with those of
various ranges of $\dot G/G$ obtained from observations as well as
from theoretical consideration, it is shown that the parameters of
the two $\Lambda$-models presented here can be tuned in most of
the cases to match with the values of $\dot G/G$ obtained from
various sources. It is worthwhile to mention here that all the
values of $\dot G/G$ listed in the Table 1 come from the
consideration that $G \sim t^{-1}$. But in the work of Milne
(1935) $G$ directly varies with $t$. Recently, Belinchon (2002),
starting from Dirac's LNH, through dimensional analysis has
arrived at the same result of Milne (1935), viz., $G \sim t$ which
obviously is in contradiction to Dirac's result. It should be
mentioned that in the present investigation also $G$ does not vary
inversely with $t$ as far as the model $\Lambda \sim \rho$ is
concerned. It has been observed by the present authors that for
$\Lambda \sim \rho$ model, $G$ varies as $t^2$. This time
variation of $G$ is different from that of Dirac (1937) and Milne
(1935), and therefore needs further investigation. Finally, it is
to be noted that the expressions for $a(t)$ and $\Lambda(t)$ for
both the models considered here maintain the same form
irrespective of the constancy or variability of $G$.

\section*{Acknowledgments}
One of the authors (SR) is thankful to the authority of
Inter-University Centre for Astronomy and Astrophysics, Pune,
India, for providing Associateship programme under which a part of
this work was carried out.




\begin{table*}
\begin{minipage}{100mm}
\caption{Values of $\alpha$ and $\beta$ for average $\dot G/G$
when
  $t_0=14$ Gyr, $\Omega_m=0.3$, $\Omega_{\Lambda}=0.7$ and $z \simeq 0$}
\label{tab1}
\begin{tabular}{@{}llrrrrlrlr@{}}
\hline Ranges of $\dot G/G$ yr$^{-1}$         &Sources &$\alpha$
&$\beta$\\ \hline $-(1.10 \pm 1.07) \times 10^{-11}<\frac{\dot
G}{G}<0$&PSR $1913 + 16 (Damour et al. 1988)$ &-0.0852&0.4074\\
\hline $-1.60 \times 10^{-12}<\frac{\dot
G}{G}<0$&Helioseismological data (Guenther et al, 1998)& -0.0115
&0.0670\\ \hline $(-1.30 \pm 2.70) \times 10^{-11}$&PSR B1855+09
(Arzoumanian 1995; Kaspi, Taylor \& Ryba 1994)&-0.1023&0.4698\\
\hline $(-8 \pm 5) \times 10^{-11}$&Lunar occultation (Van
Flandern 1975)&-1.333&1.6\\ \hline $(-6.4 \pm 2.2) \times
10^{-11}$&Lunar tidal acceleration (Van Flandern 1975)
&-0.8421&1.4328\\ \hline $-15.30 \times 10^{-11}$&Early Dirac
theory (Blake 1978)&11.7692 &2.0582\\ \hline $-5.1 \times
10^{-11}$&Additive creation theory (Blake 1978)&-0.5730&1.2644\\
\hline $(-16 \pm 11) \times 10^{-11}$&Multiplication creation
theory (Faulkner 1976)
    &8.00 &2.0869\\
\hline $-2.5 \times 10^{-10}\leq\frac{\dot G}{G}\leq+4.0 \times
10^{-11}$&WDG 117-B15A (Benvenuto et al. 2004)&-3.0 &1.8\\
\hline$\left|\frac{\dot G}{G}\right| \leq +4.10 \times
10^{-10}$&WDG 117-B15A [18]&1.1319 &3.09\\ \hline$-(0.6 \pm 4.2)
\times 10^{-12}$&Double-neutron star binaries (Thorsett
1996)&-0.0043 &0.0254\\ \hline$(0.46 \pm 1.0) \times
10^{-12}$&Lunar Laser Ranging (Turyshev et al. 2003)&0.0318
&-0.2110\\ \hline $1 \times 10^{-11 \pm 1}$&Wu and Wang
(1986)&0.0666 &-0.5\\ \hline

\end{tabular}
\end{minipage}
\end{table*}

\begin{table*}
\begin{minipage}{100mm}
\caption{Values of $\alpha$ and $\beta$ for average $\dot G/G$
when
  $t_0=6.57$ Gyr, $\Omega_m=0.3$, $\Omega_{\Lambda}=0.7$ and $z \simeq 0.5$}
\label{tab2}
\begin{tabular}{@{}llrrrrlrlr@{}}
\hline Ranges of $\dot G/G$ yr$^{-1}$         &Sources &$\alpha$
&$\beta$\\ \hline $-1.40 \times 10^{-11}<\frac{\dot G}{G}<+2.60
\times 10^{-11}$&Lor\'{e}n-Aguilar et al. (2003)&0.0196 &-0.125\\
\hline $-10^{-11} \leq \frac{\dot G}{G}<0$&Supernova Type Ia
(Gaztanaga et al. 2002) &-0.0169&0.0967\\ \hline $-4.0 \times
10^{-13}<\frac{\dot G}{G}<+3.0 \times 10^{-13}$& Big Bang
Nucleosynthesis (Copi, Davies \& Krauss 2004)&-0.0001 &0.0009\\
\hline

\end{tabular}
\end{minipage}
\end{table*}


\begin{thebibliography}{}


\bibitem{} Arzoumanian Z., 1995, Ph. D. thesis (Princeton
  University Press, Princeton, New Jersey, USA)
\bibitem{} Bartelmann M., et al., 2005, New Astron. Rev., 49, 19
\bibitem{} Belinchon J. A., 2002, Astrophys. Space Sci., 281, 765
\bibitem{} Benvenuto O. G. et al., 2004, Phys. Rev. D, 69, 082002
\bibitem{} Bertolami O. et al., 1993, Phys. Lett. B, 311, 27
\bibitem{} Biesiada M., Malec B., 2004, Mon. Not. R. Astron. Soc., 350, 644
\bibitem{} Blake G. M., 1978, Mon. Not. R. Astron. Soc., 185, 399
\bibitem{} Brans C., Dicke R. H., 1961, Phys. Lett. B, 124, 925
\bibitem{} Cetto A., Pe\~{n}a L de la., Santos E., 1986, Astron.
Astrophys., 164, 1
\bibitem{} Copi C. J., Davies A. N., Krauss L. M., 2004, Phys. Rev. Lett., 92, 171301
\bibitem{} Damour T., et al., 1988, Phys. Rev. Lett., 61, 1151
\bibitem{} Dirac P. A. M., 1937, Nature, 139, 323
\bibitem{} Dirac P. A. M., 1973, Proc. R. Soc. Lond. A, 333, 403
\bibitem{} Dyson F. J., 1972, in Aspects of quantum theory, eds Salam
A., Wigner E. P. (Cambridge Univ. Press, p. 213-216)
\bibitem{} Dyson F. J., 1978, in Current trends in the theory of fields,
eds Lannutti, Williams P. K. (American Institute of Physics, New
York, p. 163-167)
\bibitem{} Faulkner D. F., 1976, Mon. Not. R. Astron. Soc., 176, 621
\bibitem{} Gaztanaga E., et al., 2002, Phys. Rev. D, 65, 023506
\bibitem{} Gliner F., 1965, ZETF, 49, 542
\bibitem{} Goldman I., 1992, Phys. Lett. B, 281, 219
\bibitem{} Guenther D. B., et al., 1998, Ap. J., 498, 871
\bibitem{} Hellings R. W., 1987, in Problems in
  Gravitation, ed Melinkov V. (Moscow State Univ. Press, Moscow, p. 46)
\bibitem{} Hoyle F., Narlikar J. V., 1972, Mon. Not. R. Astron. Soc., 155, 323
\bibitem{} Kaspi V. M., Taylor J. H., Ryba M., 1994, Ap. J., 428, 713
\bibitem{} Knop R., et al., 2003, Astrophys. J., 598, 102
\bibitem{} Kujat J., et al., 2002, Astrophys. J., 572, 1
\bibitem{} Lor\'{e}n-Aguilar P., et al., 2003, Class. Quan.
Grav., 20, 3885
\bibitem{} Majernik V., Phys. Lett. A, 282, 362, 2001
\bibitem{} Marciano W. J., 1984, Phys. Rev. Lett., 52, 489
\bibitem{} Milne E., 1935, Relativity, gravitation and world structure (Oxford)
\bibitem{} Murphy M. T., et al., 2002, Mon. Not. R. Astron. Soc., 327, 1208
\bibitem{} Overduin J. M., Cooperstock F. I., Phys. Rev. D, 58, 043506, 1998.
\bibitem{} Perlmutter S., et al., 1999, Ap. J., 517, 565
\bibitem{} Ray S., Mukhopadhyay U., astro-ph/0407295
\bibitem{} Reasemberg R. D., 1983, Phil. Trans. R. Soc. Lond. A, 310, 227
\bibitem{} Riess A. G., et al., 1998, Astron. J., 116, 1009
\bibitem{} Stairs I. H., 2003, Living Rev. Rel., 6, 5
\bibitem{} \u{S}tefan\u{c}i{\'c} H., 2004, Phys. Lett. B, 595, 9
\bibitem{} Tegmark M., et al., 2004, Astrophys. J., 606, 70
\bibitem{} Thorsett S. E., 1996, Phys. Rev. Lett., 77, 1432
\bibitem{} Turyshev S. G., et al., 2003, gr-qc/0311039
\bibitem{} Van Flandern T. C., 1975, Mon. Not. R. Astron. Soc., 170, 333
\bibitem{} Vishwakarma R. G., 2001, Gen. Rel. Grav., 33, 1973
\bibitem{} Webb J. K., et al., 2001, Phys. Rev. Lett., 87, 091301
\bibitem{} Will C. M., 1987, in 300 Years of Gravitation (Cambridge Univ.
Press, Cambridge, p. 80)
\bibitem{} Wu Y. S., Wang Z., 1986, Phys. Rev. Lett., 57, 1978

\end{thebibliography}
\end{document}